\documentclass[12pt,preprint]{aastex}\usepackage{amssymb,epsfig}
\newcommand{\beq}{\begin{equation}}
\newcommand{\eeq}{\end{equation}}
\newcommand{\bea}{\begin{eqnarray}}
\newcommand{\eea}{\end{eqnarray}}
\newcommand{\rem}[1]{ }
\bibpunct[,]{(}{)}{;}{a}{}{,}
\begin{document}

\title{Jitter radiation as a possible mechanism for Gamma-Ray Burst
afterglows. Spectra and lightcurves} 
\author{Mikhail V. Medvedev\altaffilmark{1},}
\affil{Department of Physics and Astronomy, University of Kansas, KS 66045}
\altaffiltext{1}{
Also at the Institute for Nuclear Fusion, RRC ``Kurchatov
Institute'', Moscow 123182, Russia}
\author{Davide Lazzati, Brian C. Morsony and Jared C. Workman}
\affil{JILA, University of Colorado, 440 UCB, Boulder, CO 80309-0440}

\begin{abstract}
The standard model of GRB afterglows assumes that the radiation
observed as a delayed emission is of synchrotron origin, which
requires the shock magnetic field to be relatively homogeneous on
small scales.  An alternative mechanism -- jitter radiation, which
traditionally has been applied to the prompt emission -- substitutes
synchrotron when the magnetic field is tangled on a microscopic
scale. Such fields are produced at relativistic shocks by the
Weibel instability. Here we explore the possibility that small-scale
fields populate afterglow shocks. We derive the spectrum of jitter
radiation under the afterglow conditions. We also derive the afterglow
lightcurves for the ISM and Wind profiles of the ambient
density. Jitter self-absorption is calculated here for the first
time. We find that jitter radiation can produce afterglows similar to
synchrotron-generated ones, but with some important differences. We
compare the predictions of the two emission mechanisms. By fitting 
observational data to the synchrotron and jitter afterglow lightcurves, 
it can be possible to discriminate between the small-scale vs large-scale 
magnetic field models in afterglow shocks.  
\end{abstract}



\section{Introduction}

The general framework for the interpretation of the long-wavelength
radiation of gamma-ray burst (GRB) afterglows is the external shock
synchrotron model \citep{MR97,Wax97,P99}. In that scenario, a
blastwave is generated by the interaction of the GRB ejecta with the
interstellar medium. At the shock front, electrons are accelerated in
a power-law distribution with energy (or with Lorentz factor $\gamma$)
and a strong magnetic field is generated by some mechanism. The model
assumes that the magnetic field is coherent on the Larmor scale of the
emitting electron, hence allowing for synchrotron emission.  Both the
relativistic electron population and the magnetic field are originally
thought to share a sizable fraction of $\sim10\%$ of the internal
energy of the blastwave; those fractions are being called $\epsilon_e$
and $\epsilon_B$, respectively. Afterglow spectral fits yield typical
values, $\epsilon_e\sim0.1-0.01$ and $\epsilon_B\sim0.01-0.0001$, with
relatively large scatter \citep{PK01,P06}.

No mechanism or instability capable of generating a sub-equipartition
magnetic field in GRBs has been identified for a while, until
\citet{ML99} suggested that the field can be generated through the
Weibel instability. This prediction has been extended to
non-relativistic shocks, e.g., in supernovae and galaxy clusters
\citep{MSK05}, and confirmed via numerical PIC simulations
\citep{Silva+03,Nish+03,fred04,M+05,Spit05}.  The volume-averaged value of
$\epsilon_B$ deduced from the simulation is indeed $\sim0.01-0.0001$
(depending on location with respect to the main shock compression). 
An intriguing relation: 
\beq
\epsilon_e\simeq\epsilon_B^{1/2}, 
\eeq 
recenly found by \citet{M06b},
can allow one to reduce the number of fit parametes in afterglow studies.

Weibel-generated fields have a very short coherence length-scale
(smaller than $1/\gamma^2$ times the electron Larmor radius) and
standard synchrotron equations cannot be adopted. The theory of jitter
radiation has been proposed by \citet{M00} and further developed in
subsequent works \citep{M06,Fleish06}. (Note, however, that
\citealp{Fleish06} incorrectly predicts the absence of steep spectral
slopes, such as $F_\nu\propto\nu^1$, just below $E_{\rm peak}$ in
GRBs).

Unlike synchrotron, jitter radiation is sensitive to the statistical
properties of the magnetic field in the shock, that is to the spectrum
of magnetic fluctuations, and not just to its ``global property'' --
the strength \citep{M00}. In addition, the spectrum of jitter radiation
depends on the shock viewing angle, i.e., the angle, $\theta'$ between
the shock velocity (= propagation direction) and the line of sight
($\theta'$ is measured in the shock comoving frame).  The two extreme
cases are characterized by the emissivity function being a power-law
at frequencies below the spectral peak:
\beq
P(\omega)\propto\omega^\alpha, 
\eeq 
with $\alpha=1$ for $\theta'=0$ --
a shock viewed face-on (this is also the case of the ``effective'' 1D
magnetic turbulence, considered in \citealp{M00}), and $\alpha=0$ for
$\theta'=\pi/2$ -- an edge-on shock (this is also the case in
isotropic 2D and 3D turbulence).  These asymptotes, along with a
general case of $0\le\theta'\le\pi/2$, are considered elsewhere
(Medvedev 2006, Workman et al. in preparation).  For reference,
synchrotron radiation has $\alpha=1/3$ at low energies.

Should afterglows be resolved, one would observe a limb-brightened
object \citep{GPS99}. For a spherical outflow, the emission spectrum
will be entirely dominated by the limb emission. Because of
relativistic aberration, the photons emitted within an angle of $\sim
1/\Gamma_{\rm sh}$ with respect to the line of sight in the lab
(observer's) frame, have been emitted at angle $\theta'$ in the shock
comoving frame: 
\beq \cos\theta'=\frac{\cos(1/\Gamma_{\rm
sh})-\beta_{\rm sh}} {1-\beta_{\rm sh}\cos(1/\Gamma_{\rm sh})}= 0,
\eeq 
where $\beta_{\rm sh}$ is the dimensionless shock speed.  That is,
these photons are emitted at the angle $\theta'= 90^\circ$ away from
the shock normal. This consideration is also true for a jet outflow at
times when the jet opening angle is larger than $1/\Gamma_{\rm sh}$.

In general, after the jet break, the emission will still be dominated
by the limb. However, the shock there will not be seen edge-on, but
rather at some angle $\theta'<90^\circ$.  The jitter spectrum is, in
general, anisotropic, i.e., it depends on $\theta'$. There is a break,
$\nu_b(\theta')$ in the jitter spectrum \citep{M06} below the jitter
peak, $\nu_b<\nu_{\rm jitter~peak}$, such that $\alpha=1$ above the
break (at $\nu_b<\nu<\nu_{\rm jitter~peak}$) and $\alpha=0$ below it
(at $\nu<\nu_b$).  The dependence of the break frequency on $\theta'$
is not easy to parameterize. Since the break is quite smooth is also
not easy to establish its precise position.  One characteristic point
is: $\nu_{b} \sim 0.01 \nu_{\rm jitter~peak}$ at $\theta'=\pi/10 =
18$~degrees.

Thus, for spherical outflows and for jets seen before the ``jet
break'', the reasonable approximation will be the ``edge-on'' shock,
with $\alpha=0$. However, at late times, especially when the jet
becomes weakly relativistic, the ``face-on'' case with $\alpha=1$
should be a better approximation.

One cautious remark. The above consideration assumes strong anisotropy 
of the magnetic turbulence in the shock \citep{ML99}, which is likely 
true in the internal shocks. We do not understand well the properties
of magnetic turbulence far downstream the external shock. If the magnetic 
turbulence will become nearly isotropic, then a 3D jitter regime
gives $\alpha=0$, independent of the viewing angle.

Below, we consider both the $\alpha=0$ and the $\alpha=1$ cases in
calculating the jitter self-absorption frequency. This paper is
organized as follows: in \S~2 we summarize the shock kinematics, in
\S~3 we derive the jitter self absorption frequency in various regimes
while in \S~4 we compute all the observable properties of the
spectrum. We summarize and discuss our results in \S~5.

\section{Shock kinematics}

Here we sumarize, for future use, the kinematic relations of several
blast wave parameters. For a highly relativistic blast wave with
Lorentz factor $\Gamma_{\rm sh}$, the density jump condition relates
the pre-shock ISM density to the density downstream, $n'$, measured in
the shock frame: 
\beq 
n'\simeq4\Gamma_{\rm sh}n_{\rm ISM}.  
\eeq
The magnetic field strength downstream 
\beq 
B'=\left(32\pi\Gamma_{\rm sh}^2 n_{\rm ISM} m_p c^2\epsilon_B\right)^{1/2}, 
\eeq 
where $\epsilon_B$ is the magnetic field equipartition parameter.  The
minimum Lorentz factor of the accelerated electrons, which share a
fraction $\epsilon_e<1$ of the total energy of the shock (outflow) is:
\beq 
\gamma_m=\left(\frac{s-2}{s-1}\right)\frac{m_p}{m_e}\
\epsilon_e\Gamma_{\rm sh}\approx6.12\times10^2\epsilon_e
\Gamma_{\rm sh}.  
\eeq 
Here the last expression is calculated for a typical electron
power-law index $s=2.5$, see definition below, Eq (\ref{N}).

\subsection{Constant density ISM}

The radius and Lorentz factor of a blast wave propagating in a
constant density ISM depend on the observed time as \citep{GPS99}
\begin{eqnarray}
\Gamma_{\rm sh}&\approx& 3.65\left(\frac{E_{52}}{n_{\rm
ISM,0}}\right)^{1/8} \left(\frac{t_{\rm days}}{1+z}\right)^{-3/8}, \\
R&\approx& 5.53\times10^{17} \left[\frac{E_{52} t_{\rm days}}{n_{\rm
ISM,0}(1+z)} \right]^{1/4}\textrm{ cm}.
\label{ISM}
\end{eqnarray}
Here $E_{52}=E_{\rm explosion}/10^{52}$~erg, and $n_{\rm ISM,0}=n_{\rm
ISM}/1$~cm$^{-3}$.

\subsection{Wind model}

If the blast wave is propagating in the wind environment with the
density dicreasing with distance, $n\propto r^{-2}$, the Lorentz
factor and the radius of the blast wave are \citep{CL00}
\begin{eqnarray}
\Gamma_{\rm sh}&\approx& 4.96\,
E_{52}^{1/2}\,\left(\frac{A_* t_{\rm days}}{1+z}\right)^{-1/4}, \\
R&\approx& 1.56\times10^{17}
\left[\frac{ E_{52} t_{\rm days}}{A_* (1+z)}
\right]^{1/2}\textrm{ cm}.
\label{Wind}
\end{eqnarray}
Here the wind parameter $A_*=[\dot M_w/(10^{-5}M_{\sun}\textrm{
yr}^{-1})]/ [V_w/(10^3\textrm{ km s}^{-1})]$, where $\dot M_w$ is the
mass loss rate, $V_w$ is the wind velocity. Since the ambient density
is no longer constant, one need to substitute $n_{\rm ISM}$ with the
wind density 
\beq 
n_{\rm wind}=AR^{-2}\approx(3.00\times10^{35}\,A_*\textrm{ cm}^{-1})\,R^{-2}.
\eeq

\section{Theory}

Here we work in the shock comoving frame. Thus, all the frequencies
are expressed in this frame. Also, all the shock parameters, such as
the particle number density $n'$ and magnetic field $B'$, are those in
the shocked region and are measured in the comoving frame as well,
unless stated otherwise. We use ``prime'' to denote quantities
measured in the shock frame. Sometimes we omit ``prime'' when this
does not cause any confusion.

We introduce the accelerated electron distribution function:
\beq
N(\gamma)=(s-1)N_e\gamma_m^{s-1}\gamma^{-s}, \qquad \gamma\ge\gamma_m,
\label{N}
\eeq
where $s$ is the power-law index, $\gamma_m$ is the minimum Lorenz
factor (low-energy cut-off), $N_e=4\pi R^2\Delta' n_{\rm ISM}$ is the
total number of the nonthermal (emitting) electrons, $R$ is the radius
of the blast wave, $\Delta'$ is its thickness, $n_{\rm ISM}$ is the
number density of the electrons in the ambient medium.

The absorption coefficient at comovong frequency $\nu'$ is
\citep{DBC00}:
\beq
\kappa_{\nu'}=\frac{1}{8\pi m_e V_{\rm bw} \nu'^2}
\int_1^{\infty} d\gamma\; P(\nu',\gamma)~
\gamma^2\frac{\partial}{\partial\gamma}\left[\frac{N(\gamma)}{\gamma^2}\right],
\eeq
where $V_{\rm bw}=4\pi r^2\Delta'$ is the volume of the blast wave of
the comoving thickness $\Delta'$ and $ P(\nu',\gamma)$ is the
emissivity function. Straightforwardly, 
\beq
\gamma^2\frac{\partial}{\partial\gamma}\left[\frac{N(\gamma)}{\gamma^2}\right]
=-(s+2)(s-1)N_e\gamma_m^{s-1}\gamma^{-s-1}.
\label{kappa}
\eeq

\subsection{Emissivity functions}

We first discuss synchrotron, for reference.  The synchrotron comoving
peak frequency is at 
\beq \nu'_s=(3/2)\nu_B\gamma^2, 
\eeq 
where $\nu_B=eB'/2\pi m_ec$ is the cyclotron (Larmor) frequency in a
homogeneous magnetic field of strength $B'$, the pitch-angle is
assumed to be $\pi/2$, for simplicity.  At $\nu'\ll\nu'_s$, the
synchrotron emissivity is 
\beq P_{\rm synch}(\nu',\gamma)\simeq
\frac{e^2}{c}\sqrt{3}\nu_B\left[\frac{4\pi}
{\sqrt{3}\Gamma(1/3)}\left(\frac{\nu'}{3\nu_B\gamma^2}\right)\right],
\eeq 
where $\Gamma(1/3)\simeq2.68$ is a gamma function.

The jitter peak frequency is determined by the magnetic field spectrum
in the post-shock medium.  Recent numerical simulations
\citep{fred04,MSK05} demonstrate that the field generation by both the
electrons and the protons, occurs well before the main shock
compression. The wave-vector of the fastest growing mode of the Weibel
instability in the linear regime is \citep{ML99}: 
\beq 
k_{\rm Weibel}\approx 2^{-1/4}(\omega_{p,e}/c){\bar\gamma_e}^{-1/2}, 
\eeq
where $\omega_{p,e}=\sqrt{4\pi e^2 n/m_e}\approx
5.64\times10^4n^{1/2}$~s$^{-1}$ is the plasma frequency. Because
$\omega_{p,e}$ is a Lorentz invariant, $k_{\rm Weibel}$ is determined
by the parameters of the ambient, unshocked medium alone. Thus, the
density $n=n_{\rm ISM}$ is the ambient medium density, and the mean
Lorentz factor of the electrons in the ambient medium, $\bar\gamma_e$,
is close unity (the ISM is cold, non-relativistic). The shock
compression and the proton thermalization occur far downstream,
significantly behind the region where the field has been
generated. Therefore, the correlation length of the field decreases
(primarily in the parallel direction) because of the
compression. Hence, the characteristic wave-vector of the downstream
random fields is
\beq 
k_{\rm rand}\simeq (4\Gamma_{\rm sh})k_{\rm  Weibel}.  
\eeq 
Numerical simulations also indicate that the correlation scale of the
field $\lambda_B=2\pi/k_{\rm rand}$ varies with distance from the
shock front because of the highly nonlinear dynamics of the
Weibel-generated currents and fields \citep{fred04,M+05}. We therefore
parameterize this as 
\beq 
k_{\rm rand}=\eta(\omega_{p,e}/c), 
\eeq
where the parameter $\eta$ incorporates all relativistic effects, the
shock compression and the nonlinear evolution of the Weibel
turbulence.  Thereafter, we use 
\beq 
\eta\simeq 2^{-1/4}(4\Gamma_{\rm sh})\bar\gamma_e^{-1/2}
\simeq2^{7/4}\Gamma_{\rm sh}, 
\eeq 
where we assumed that $\bar\gamma_e\sim1$.

The characteristic frequency of the electron's jitter while it moves
at roughly the speed of light through these magentic fields is
$\nu\sim c/\lambda_B$. More precisely, 
\beq 
\nu_r=k_{\rm rand}c/2\pi=\eta\omega_{p,e}/2\pi=\eta\nu_{p,{\rm ISM}}, 
\eeq 
where we introduced the plasma frequency of the ISM, $\nu_{p,{\rm
ISM}}=(4\pi e^2 n_{\rm ISM}/m_e)^{1/2}/2\pi =8.98\times10^3 n_{\rm
ISM}^{1/2}$~Hz.

The peak of the emitted jitter radiation is $\nu'_j\sim\nu_r\gamma^2$,
but its exact position slightly depends on the magnetic field
spectrum, defined as $(B'^2)_k\propto k^{2\mu}$. For steep spectra,
$\mu\gg1$, the peak is at, roughly, twice the jitter frequency: 
\beq
\nu'_{j}\simeq 2\nu_r\gamma^2. 
\eeq 
The jitter and synchrotron peak frequencies in the magnetic fields of
identical strengths are related to each other via the identities \beq
\nu_r=\nu_B/\delta, \textrm{ or } \nu'_{j}=\nu'_s/(3\delta/4), \eeq
These identities define the parameter $\delta\lesssim1$, which is the
ratio of the deflection angle of the particle path in chaotic fields
to the beaming angle $1/\gamma$ \citep{M00}. It is expressed via the
magnetic field equipartition parameter in the shock, $\epsilon_B$, as
\beq \delta=\left(\frac{m_p}{m_e}\frac{8\Gamma_{\rm sh}}{\eta^2}\,
{\epsilon_B}\right)^{1/2} \simeq
\left(\frac{m_p}{m_e}\,2^{-1/2}{\epsilon_B}\right)^{1/2} \approx
36.0\sqrt{\epsilon_B}.  \eeq The jitter emissivity function below the
peak is \beq P_{\rm jitter}(\nu',\gamma)\simeq\frac{e^2}{c}\pi
f(\mu)\delta^2\nu_r \left(\frac{\nu'}{2\nu_r\gamma^2}\right)^\alpha.
\eeq Later on, we neglected a factor $f(\mu)=(2\mu+1)/(2\mu-1)$
(calculated for $\alpha=1$) because it is of order unity when
$\mu\gg1$.

Thus, for both emission mechanisms,
\beq
P(\nu',\gamma)\simeq a \left(\frac{\nu'}{b\gamma^2}\right)^\alpha.
\label{P}
\eeq
Here, for synchrotron
\beq
a=\frac{e^2}{c}\frac{4\pi}{\Gamma(1/3)}\;\nu_B, \qquad
b=3\nu_B, \qquad
\alpha=1/3,
\eeq
and for jitter 
\beq
a=\frac{e^2}{2c}\pi\delta^2\nu_r,\qquad
b=\nu_r,\qquad
\alpha=0,\textrm{  and  } \alpha=1.
\eeq

\subsection{Self-absorption frequencies for $\nu_a<\nu_{m}$}

Here we assume that the self-absorption frequency is below the jitter
peak in the spectrum from the ensemble of the electrons,
$\nu'_a\la\nu'_{m}=2\nu_{j,m}\equiv 2\nu_r\gamma_m^2$.

The opacity is easily calculated from Eqs. (\ref{kappa}), (\ref{P}) to
yield 
\beq 
\kappa_{\nu'}=\frac{(s+2)(s-1)N_e\; a\;
b^{-\alpha}\;(\nu')^{\alpha-2}} {8\pi m_e V_{\rm bw} (s+2\alpha)
\gamma_m^{2\alpha+1}}.  
\eeq

The self-absorption frequency is that, at which the optical thickness
of the blast wave shell is unity:
\beq
\kappa_{\nu'_a}\Delta'\sim1.
\eeq
This condition gives:
\beq
\nu'^{2-\alpha}_a\simeq\frac{(s+2)(s-1)}{(s+2\alpha)}
\frac{R\; n_{\rm ISM}\; a\; b^{-\alpha}}{24\pi m_e\gamma_m^{2\alpha+1}}. 
\eeq
This equation works for both synchrotron and jitter radiation.  One
just need to replace the general parameters $a,\ b,\ \alpha$ with
those above, for a process of interest.

For jitter radiation, we have
\beq
a\;b^{-\alpha}=\frac{e^2}{c}\frac{\pi}{2^\alpha}\delta^2\nu_r^{1-\alpha}.
\eeq
Therefore,
\beq
\nu'^{2-\alpha}_a\simeq\frac{(s+2)(s-1)}{(s+2\alpha)}
\frac{\pi}{24}
\frac{(R/c) \nu_{p,{\rm ISM}}^2\nu_r^{1-\alpha}\delta^2}
{\gamma_m^{2\alpha+1}}. 
\eeq

\subsubsection{Case $\alpha=0$}

In this case,
\begin{eqnarray}
\nu'_a&\simeq&\left(\frac{\pi}{24}\frac{(s+2)(s-1)}{s}\right)^{1/2}
\frac{(R/c)^{1/2}\nu_{p,{\rm ISM}}\nu_r^{1/2}\delta}{\gamma_m^{1/2}} 
\nonumber\\
&\approx&2.92\textrm{ Hz } 
R^{1/2}n_{\rm ISM}^{3/4}\gamma_m^{-1/2}\eta^{1/2}\delta,
\nonumber\\
&\approx&5.36\textrm{ Hz } 
R^{1/2}n_{\rm ISM}^{3/4}\gamma_m^{-1/2}\Gamma_{\rm sh}^{1/2}\delta.
\end{eqnarray}
Hereafter, we use a typical value, $s=2.5$,  in numerical estimates.

\subsubsection{Case $\alpha=1$}

This case may be relevant to the late afterglow from a jet. 
\begin{eqnarray}
\nu'_a&\simeq&(s-1)\frac{\pi}{24}
\frac{(R/c)\nu_{p,{\rm ISM}}^{2}\delta^2}{\gamma_m^{3}} 
\nonumber\\
&\approx&5.28\times10^{-4}\textrm{ Hz } R\,n_{\rm ISM}\,\gamma_m^{-3}\delta^2.
\end{eqnarray}

\subsection{Self-absorption frequency for $\nu_a>\nu_m$}

In order to calculate the self-absorption frequency in this regime, 
we need the full expression for the emissivity function:
\beq
P(\nu',\gamma)
=\frac{e^2}{2c}\,\delta\,2\pi\nu_r\,
J\left(\frac{\nu'}{\nu_r\gamma^2}\right), 
\eeq
where the function $J(\xi)$ for $\mu\gg1$ and $\delta\ll1$ is
\begin{eqnarray}
J(\xi)&=&(2\mu+1)\xi^{2\mu}\left[I(2)-I(\xi)\right], \\
I(\xi)&=&-\left(\frac{\xi^{-2\mu+1}}{2\mu-1}-
\frac{\xi^{-2\mu+2}}{2\mu-2}+\frac{1}{2}\frac{\xi^{-2\mu+3}}{2\mu-3}\right).
\end{eqnarray}
This function corresponds to the $\alpha=1$ case. The expression for
the $\alpha=0$ is more complicated. Our analysis indicates, however,
that the jitter emissivity function is approximated by a sharply
broken power-law very well (much better than the synchrotron one does).
Therefore, we will use the approximate expression for $J(\xi)$,
which describes well the $\alpha=1$ and $\alpha=0$ spectra.
We use
\beq
J(\xi)=\left\{
\begin{array}{ll}
\xi^\alpha, & \textrm{  if } \xi<2,\\
0, & \textrm{  if } \xi\ge2.
\end{array}\right.
\eeq
This function mimics a power-law spectrum up to the
peak jitter frequency $\nu_{j}\simeq 2\nu_r\gamma^2$,
with the sharp cutoff above it.

The opacity is:
\begin{eqnarray}
\kappa_{\nu'}=
\frac{(s+2)(s-1)N_e\gamma_m^{s-1}}{8\pi m_e V_{\rm bw}\nu'^2}
\frac{e^2}{2c}\delta 2\pi\nu_r \int_{\gamma_m}^\infty 
J\left(\frac{\nu'}{\nu_r\gamma^2}\right)\gamma^{-s-1}\, d\gamma.
\end{eqnarray}
Upon substitution of $J(\xi)$ the integral becomes
\beq
\frac{1}{2}\left(\frac{\nu}{\nu_r}\right)^{-s/2}
\int_0^{\textrm{min}\!\left({\nu'}/({\nu_r\gamma_m^2});~ 2\right)}
z^{\alpha+{s}/{2}-1}\; dz.
\eeq
We are interested in the high-energy part of the spectrum, above
$\nu'\gg\nu'_m\simeq2\nu_r\gamma_m^2$. Therefore, the upper limit in
the integral is equal to 2. Thus, we have for the opacity at
frequencies above the spectral peak:
\beq
\kappa_{\nu'}=\frac{(s+2)(s-1)}{(s+2\alpha)}\frac{2^{s/2+\alpha-3}}{12\pi}
\frac{(R/c)\omega_{p,e,{\rm ISM}}^2}{\Delta'\gamma_m^3}\frac{\delta^2}{\nu'_m}
\left(\frac{\nu'}{\nu'_m}\right)^{-s/2-2}.
\eeq
Since $F_\nu\propto\nu^{5/2}$ for $\nu_m<\nu<\nu_a$, rather than
$\propto\nu^2$, the absorption frequency should be defined as the
frequency at which $dF_\nu/d\nu=0$, where
$F_\nu\propto\nu^{5/2}(1-e^{-\kappa_{\nu'}\Delta'})$.  For a simple
estimate, we can still use the condition:
$\kappa_{\nu'_a}\Delta'\simeq1$. We have 
\beq 
\nu'_a\simeq\left(
2^{s/2+\alpha-3}\frac{\pi}{3}\frac{(s+2)(s-1)}{(s+2\alpha)}\right)^{2/(s+4)}
\left[\frac{(R/c)\nu_{p,{\rm
ISM}}^2\nu'^{s/2+1}_m\delta^2}{\gamma_m^3} \right]^{2/(s+4)}.
\label{nu-a39}
\eeq
The peak frequency, $\nu'_m$, is calculated in the next section, 
Eq. (\ref{nu-m}). To proceed furter, we again use the typical value of 
the electron power-law exponent, $s=2.5$. For such $s$, the 
values of the numerical factor (the first term) in Eq. (\ref{nu-a39}) is
equal to 0.95 and 0.99 for $\alpha=0$ and $\alpha=1$, respectively. 
We, therefore
use the representative value of 0.97 for both cases, which gives
an error in the $\nu'_a$ value to within a couple of percents. Thus,
\beq
\nu'_a\approx320\textrm{ Hz } R^{0.31}n_{\rm ISM}^{0.65}
\Gamma_{\rm sh}^{0.69}\gamma_m^{0.46}\delta^{0.62}.
\eeq

\subsection{Peak frequency}

The peak frequency of the shock spectrum (from an ensemble of electrons)
is determined by the jitter frequency at $\gamma_m$:
\begin{eqnarray}
\nu'_m&\simeq&2\nu_r\gamma_m^2 \nonumber\\
&\approx&1.80\times10^4\textrm{ Hz } n_{\rm ISM}^{1/2}\gamma_m^2\eta
\nonumber\\ 
&\approx&6.04\times10^4\textrm{ Hz } n_{\rm
ISM}^{1/2}\Gamma_{\rm sh}\gamma_m^2.
\label{nu-m}
\end{eqnarray}

\subsection{Cooling break frequency}

The total (integrated over frequencies) emitted power by an electron
is identical in jitter and synchrotron regimes.  Thus, the cooling
break is unchanged. We quote results from \citet{SPN98} 
\beq
\nu'_c=(3/2)\nu_B\gamma_c^2, 
\eeq 
where the cooling Lorentz factor (neglecting Compton losses) is 
\beq
\gamma_c=\frac{3m_e}{16m_pc\sigma_T\epsilon_B} \frac{1}{n_{\rm
ISM}\Gamma_{\rm sh}t_{\rm loc}} 
\eeq 
where $t_{\rm loc}=t/(1+z)$ is the cosmologically local time for a
GRB. Thus, 
\beq 
\nu'_c\approx4.29\times10^{24}\textrm{ Hz } n_{\rm
ISM}^{-3/2} \Gamma_{\rm sh}^{-5}\epsilon_B^{-3/2} t_{\rm loc}^{-2}.
\eeq

\section{Observables}

The power-law segments in $\nu_a<\nu_m$ and $\nu_a>\nu_m$ regimes are, 
respectively:
\beq
F_\nu^{(\nu_a<\nu_m)}\propto\left\{
\begin{array}{ll}
\nu^2, & \textrm{ if   } \nu<\nu_a; \\
\nu^\alpha, & \textrm{ if   } \nu_a<\nu<\nu_m; \\
\nu^{-(s-1)/2}, & \textrm{ if   } \nu_m<\nu<\nu_{c}; \\
\nu^{-s/2}, & \textrm{ if   } \nu_{c}<\nu.
\end{array}
\right.
\eeq
\beq
F_\nu^{(\nu_a>\nu_m)}\propto\left\{
\begin{array}{ll}
\nu^2, & \textrm{ if   } \nu<\nu_m; \\
\nu^{5/2}, & \textrm{ if   } \nu_m<\nu<\nu_a; \\
\nu^{-(s-1)/2}, & \textrm{ if   } \nu_a<\nu<\nu_{c}; \\
\nu^{-s/2}, & \textrm{ if   } \nu_{c}<\nu.
\end{array}
\right.
\eeq

\subsection{Frequencies: ISM}

We now calculate the frequencies in the observer's frame 
(thus, all frequencies are boosted by $\Gamma_{\rm sh}/(1+z)$) and their
dependencies on the burst parameters for $s=2.5$.
\begin{eqnarray}
\nu_a^{(\alpha=0)}&\approx& 5.88\times10^{8}\textrm{ Hz }
(1+z)^{-3/4}E_{52}^{1/4}\epsilon_e^{-1/2}\delta\,
n_{\rm ISM,0}^{1/2} t_{\rm days}^{-1/4}, 
\\
\nu_a^{(\alpha=1)}&\approx& 9.56\times10^{4}\textrm{ Hz }
(1+z)^{-2}E_{52}^{0}\epsilon_e^{-3}\delta^2
n_{\rm ISM,0}^{} t_{\rm days}^{},
\\
\nu_a^{(>\nu_m)}&\approx& 3.13\times10^{10}\textrm{ Hz }
(1+z)^{-0.27} E_{52}^{0.35}\epsilon_e^{0.46}\delta^{0.62}
n_{\rm ISM,0}^{0.30}t_{\rm days}^{-0.73},
\\
\nu_m&\approx& 4.02\times10^{12}\textrm{ Hz }
(1+z)^{1/2}E_{52}^{1/2}\epsilon_e^2 t_{\rm days}^{-3/2},
\\
\nu_c&\approx& 8.54\times10^{16}\textrm{ Hz } (1+z)^{-1/2}
E_{52}^{-1/2}\epsilon_{B,-3}^{-3/2}n_{\rm ISM,0}^{-1}t_{\rm days}^{-1/2}.
\end{eqnarray}
Here we also quoted the result from \citet{SPN98} for the
cooling frequency in the adiabatic regime of blast wave evolution
in the constant ISM density environment,
corrected for the redshift.

Finally, we compute the peak flux of jitter afterglows by using the
relation $F_{\nu,\max}=\delta^2 F_{\nu,\max,{\rm synch}}
\nu_m/\nu_{m,{\rm synch}}$ and the synchrotron peak flux from Sari et
al. (1998):
\begin{equation}
F_{\nu,\max} = 10^3 \,E_{52}\,\epsilon_{B,-3}\,n_{\rm{ISM},0}\,
D_{28}^{-2}\,\mu Jy
\end{equation}

We remind that the jitter deflection parameter is related 
to the magnetic field equipartition parameter as follows:
\beq
\delta\approx36.0\epsilon_B^{1/2}\approx1.12\epsilon_{B,-3}^{1/2}.
\eeq

\begin{figure}
\psfig{file=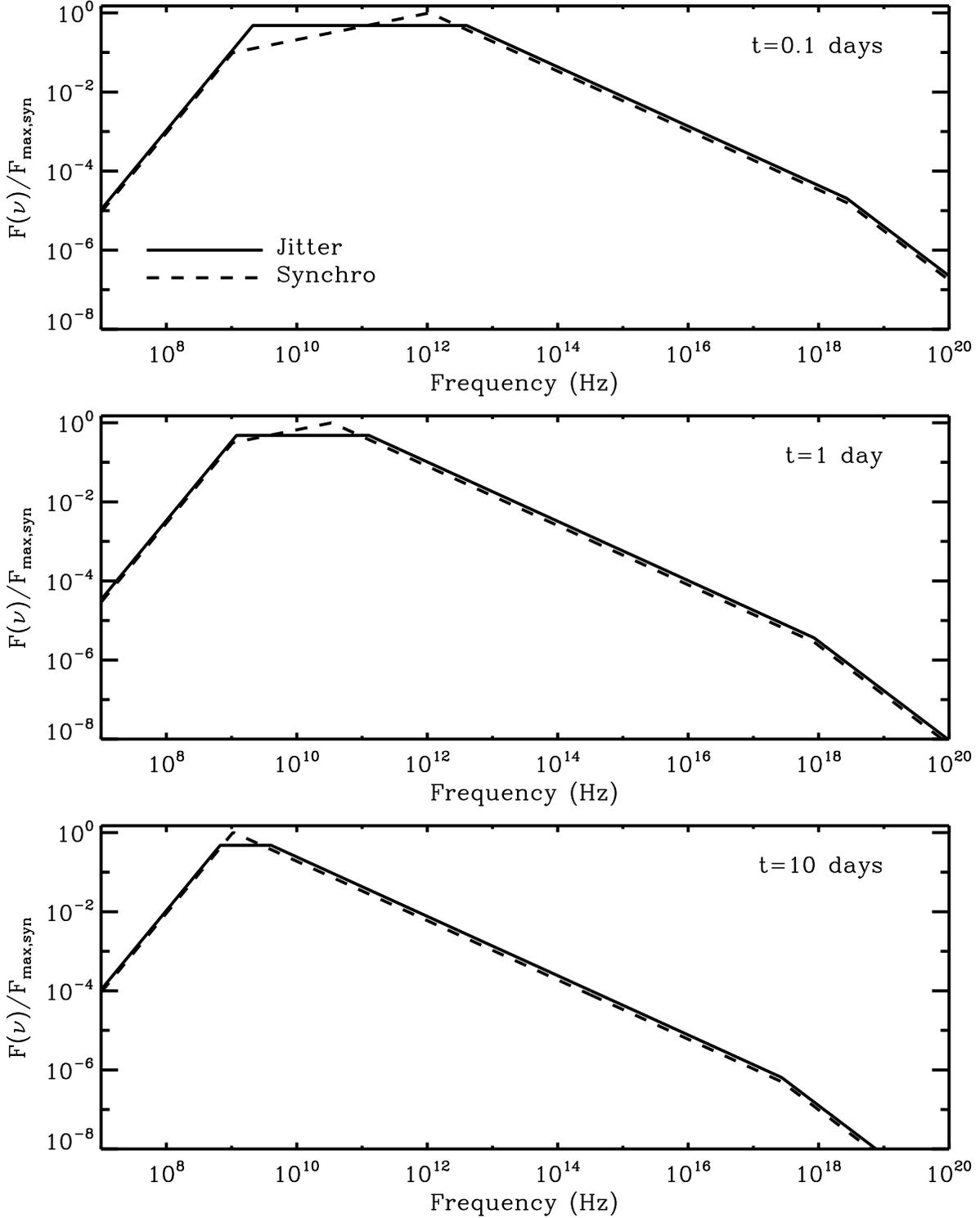}
\caption{{Jitter and synchrotron spectra for a typical afterglow
running into a uniform medium. The parameter set is: $E=10^{53}$~erg,
$\epsilon_e=0.1$, $\epsilon_B=0.0001$, $n_{\rm ISM}=1$ and the electron
energy distribution power-law index $s=2.5$,
computed at $t=$0.1, 1 and 10 days.}
\label{fig:spex}}
\end{figure}

Figure~\ref{fig:spex} shows the jitter and synchrotron spectra for an
afterglow with $E=10^{53}$~erg, $\epsilon_e=0.1$, $\epsilon_B=0.0001$,
$n_{ISM}=1$ and $s=2.5$, computed at $t=$0.1, 1 and 10 days.

\subsection{Frequencies: Wind}

We now calculate the frequencies in the observer's frame 
in the wind environment.
\begin{eqnarray}
\nu_a^{(\alpha=0)}&\approx& 2.79\times10^{9}\textrm{ Hz }
(1+z)^{-1/4}E_{52}^{0}\epsilon_e^{-1/2}\delta\,
A_*^{} t_{\rm days}^{-3/4}, 
\\
\nu_a^{(\alpha=1)}&\approx& 1.80\times10^{5}\textrm{ Hz }
(1+z)^{-1}E_{52}^{-3/2}\epsilon_e^{-3}\delta^2
A_*^{2} t_{\rm days}^{0},
\\
\nu_a^{(>\nu_m)}&\approx& 1.41\times10^{11}\textrm{ Hz }
(1+z)^{0.04} E_{52}^{0.58}\epsilon_e^{0.46}\delta^{0.62}
A_*^{0.62}t_{\rm days}^{-1.04},
\\
\nu_m&\approx& 2.40\times10^{13}\textrm{ Hz }
(1+z)^{1/2}E_{52}^{3/2}\epsilon_e^2 t_{\rm days}^{-3/2},
\\
\nu_c&\approx& 5.66\times10^{15}\textrm{ Hz } (1+z)^{-3/2}
E_{52}^{1/2}\epsilon_{B,-3}^{3/2}A_*^{-2}t_{\rm days}^{1/2}.
\end{eqnarray}
Here we also quoted the result from \citet{CL00} for the
cooling frequency in the adiabatic regime of blast wave evolution
in the wind environment. 

\section{Discussion}

We considered in this work the propertiies of GRB afterglows with
radiation produced by jitter radiation instead of synchrotron. For the
first time we evaluate the self absorption frequency in various
regimes and for blastwaves rpopagating in different ambient media.

Within the present framework, we analyzed two possible regimes of the
jitter mechanism. If the post-shock magnetic turbulence is isotropic
(which is very likely in the far downstream region), then
$\alpha=0$. If the turbulence remains anisotropic throughout the shell
than (i) if we observe a shock before the jet break or if the outflow
is spherical, then still $\alpha=0$. However, (ii) for anisotropic
turbulence and late times (long after the jet break), one can have
$\alpha=1$. It is likely that one can have an intermediate regime for
most of the time, as the properties of turbulence likely vary
downstream as a function of the distance from the shock front.

Note that in the jitter regime, the peak frequency is independent of
the magnetic field strength. In general, it depends on the ambient
density. However, for the assumed parameter $\eta\propto\Gamma_{\rm
sh}$, which incorporates all details (not so well known) of the
magnetic field evolution far downstream, the jitter peak is
independent of the density at all, either $n_{\rm ISM}$ or $A_*$.
Note also that the jitter self-absorption frequencies and the peak
frequency strongly depend on the electron equipartition parameter.  It
may be helpful to remember that the jitter peak frequency is higher
than the synchrotron peak frequency in the field of the same strength
(same $\epsilon_B$) and the same electron energy distribution (same
$\epsilon_e$ and the electron index $s$) by a factor $\sim\delta^{-1}$, 
which, in turn, is $\propto\epsilon_B^{1/2}$.

Finally, we estimate the times when the emission is becomes optically 
thick, $\nu_a\sim\nu_m$ for both $\alpha$'s, for the ISM case.
\begin{eqnarray}
t_a^{(\alpha=0)}&\sim&1170\textrm{ days }(1+z)E_{52}^{1/5}
\epsilon_e^{2}\delta^{-4/5}n_{\rm ISM,0}^{-2/5}, \\
t_a^{(\alpha=1)}&\sim& 1120\textrm{ days }(1+z)E_{52}^{1/5}
\epsilon_e^{2}\delta^{-4/5}n_{\rm ISM,0}^{-2/5},\\
t_a^{(\nu_m<\nu_a)}&\sim& 548\textrm{ days }(1+z)E_{52}^{0.20}
\epsilon_e^{2.00}\delta^{-0.80}n_{\rm ISM,0}^{0.40}
\end{eqnarray}
Ideally, these times should coincide. The discrepancies are due to the
approximations made in our analysis. In particular, the
self-absorption frequency for $\nu_m<\nu_a$ is overestimated by a
factor of two (as is explained above), hence the thin-to-thick
transition time is earlier. A more detailed treatment of the
self-absorption frequencies will be presented elsewere (Jared et
al. in preparation; Morsony et al. in preparation).  Also, the time when
the cooling break is equal to the peak, $\nu_c\sim\nu_c$ is 
\beq
t_c\sim4.71\times10^{-5} \textrm{ days }
(1+z)E_{52}\epsilon_e^2\epsilon_{B,-3}^{3/2}n_{\rm ISM,0}.  
\eeq 
The numerical factor in the above equation is equal to 4.07 seconds.

Similartly, we calculate $t_a$'s and $t_c$ for the wind case. 
\begin{eqnarray}
t_a^{(\alpha=0)}&\sim&1.76\times10^{5}\textrm{ days }(1+z)E_{52}^{2}
\epsilon_e^{10/3}\delta^{-4/3}A_*^{-4/3}, \\
t_a^{(\alpha=1)}&\sim& 2.61\times10^{5}\textrm{ days }(1+z)E_{52}^{2}
\epsilon_e^{10/3}\delta^{-4/3}A_*^{-4/3}, \\
t_a^{(\nu_m<\nu_a)}&\sim& 7.08\times10^{4}\textrm{ days }(1+z)E_{52}^{2.00}
\epsilon_e^{-3.35}\delta^{-1.34}A_*^{-1.34}, \\
t_c&\sim&6.51\times10^{-2} \textrm{ days }
(1+z)E_{52}^{1/2}\epsilon_e^{}\epsilon_{B,-3}^{-3/4}A_*.
\end{eqnarray}
The discrepancies in $t_a$'s are, again, due to the approximations
made in calculating $\nu_a$.

Figure~\ref{fig:spex} allows us to comment on the differences between
synchrotron and jitter afterglows. The high energy part of the
spectrum (mainly the optical and X-ray regimes) are hardly
distinguishable between the two mechanisms, especially if, as expected
from simulation, jitter fields are created such that
$\delta\lesssim1$.

The two main differences are in the low energy branches, around when
most radio observations are performed. First, the spectral slope at
the left of the peak frequency is flat, $\propto\nu^0$, rather than
the canonical $\nu^{1/3}$; second, the location of the self-absorption
break is different and evolves in time differently than in synchrotron
afterglows. Since one important observation to nail down the density
of the ambient medium is the radio regime, modelling afterglows with
jitter radiation may lead to different results compared to those of
synchtron modelling (Morsony et al. and Workman et al., in
preparation).

\acknowledgements

MM gratefully acknowledges support from 
the Institute for Advanced Study.
This work was supported by NASA grants NNG-04GM41G (MM) and
NNG-06GI06G (BM, DL), Swift Guest Investigator grant 06-SWIFT306-0001
(MM) and NNX06AB69G (BM, DL), DoE grant DE-FG02-04ER54790 (MM), and
NSF grant AST-0307502 (BM, DL).


\rem{
%
\begin{figure}
\caption{ 
\label{f:1} }
\end{figure} 
}

\end{document}